\documentclass[prl,aps,twocolumns]{revtex4}
\usepackage{amsmath,bm}
\usepackage{graphicx}

\begin{document}

\title{Topological thermal instability and the length of proteins}

\author{Raffaella Burioni}
\affiliation{Dipartimento di Fisica and INFM, Universit\`a di Parma, 
Parco Area delle Scienze 7A, 43100 Parma, Italy}
\author{Davide Cassi}
\affiliation{Dipartimento di Fisica and INFM, Universit\`a di Parma, 
Parco Area delle Scienze 7A, 43100 Parma, Italy}
\author{Fabio Cecconi}
\affiliation{Dipartimento di Fisica and INFM (UdR and SMC) , Universit\`a
di Roma La Sapienza
Piazzale A. Moro 2, 00185 Roma, Italy}
\author{Angelo Vulpiani}
\affiliation{Dipartimento di Fisica and INFM (UdR and SMC) , Universit\`a
di Roma La Sapienza
Piazzale A. Moro 2, 00185 Roma, Italy}

\begin{abstract}
We present an analysis of the role of
global topology on the structural stability of folded proteins in thermal
equilibrium with a heat bath. For a large class of single domain proteins, we
compute the harmonic spectrum within the Gaussian Network Model (GNM) and we
determine the spectral dimension, a parameter describing the low frequency
behaviour of the density of modes. 
We find a surprisingly strong correlation
between the spectral dimension and the number of aminoacids of the protein. 
Considering
that the larger the spectral dimension, the more topologically compact is the
folded state, our results indicate that for a given temperature and length of
the protein, the folded structure corresponds to the less compact folding
compatible with thermodynamic stability.
\vskip 1 truecm

\end{abstract}
\maketitle

\section{INTRODUCTION}
The role of geometry has recently been considered as a factor of primary 
importance for the comprehension of several physical properties of 
proteins 
and other biological macromolecules. In particular, 
since the topology of folded states is known to influence the folding 
properties
\cite{Go,Plaxco,Makarov,Baker,Riddle,Clementi,Thirumalai,Cecconi}
, a great deal of work 
has been devoted to the study of topological aspects describing the 
networks of links between aminoacids in folded proteins. 
\cite{Kaba,Park,Vendru}. 
Futhermore, relevant features of protein conformations seem to follow 
the geometrical principles of the optimal packing 
problem~\cite{Maritan,Banavar}.   

Starting from the primary, linear, structure (the
sequence of aminoacids), a protein evolves during the folding process 
until it reaches a final state (native state) whose geometrical shape is 
crucial to the function of the protein itself. However, the problem of the 
topological arrangement of proteins in their native states cannot be 
regarded as purely static issue. Indeed, a massive accumulation of 
experimental data collected from X-ray, NMR and neutron spectroscopy,
has reveiled that protein native states are rather 
dynamic structures where aminoacids constantly move around 
equilibrium positions. 
This dynamics, crucially involved in protein 
functions \cite{Frauen,Frauen2}, is usually examined and 
investigated through normal modes analysis (NMA) \cite{NMA} and its 
variants \cite{Essential}. 

The study of collective motions of 
large proteins is generally of difficult accessibility to realistic all-atoms 
NMA~\cite{Levitt} and simplified or approximate approaches are usually 
welcome. 
Tirion \cite{Tirion} first
proposed the possibility to replace, in protein normal mode computations, 
complicated empirical potentials by Hookian pairwise interactions depending 
on a single parameter. 
The approach stems from the observation that low-frequency dynamics,
which is mainly associated with protein-domain motion, is generally
insensitive to the finer details of atomic interactions.
A vast further literature \cite{Hinsen,Bahar,Atilgan,Haliloglu} confirmed
the success of simple harmonic models in the study of slow vibrational 
dynamics of large biological macromolecules, until they become 
a viable alternative to heavy and time-consuming all-atoms NMA. 
This success relies on the striking agreement with experiments, on the 
presence of few adjustable parameters and finally on their easy numerical 
implementation on computers and fast result production.
For such reasons harmonic models are suitable also for the
systematic analysis of large datasets of proteins.

The topological stability of macromolecules is far from being
a pure mechanical problem as it closely involves thermodynamics: indeed,
the relevant thermodynamic potential to minimize in order to find 
the stable configuration is not an energy, 
but a free energy. This is due to the interaction with
the environment (schematized as a thermal bath) which is generally not  
negligible, specifically for biological macromolecules having a stable 
phase in a solvent. In particular, water is a very efficient medium 
for the transfer of thermal energy at microscopic scales 
(i.e. oscillations and molecular rotations).

Following these considerations, in this work we apply the approach 
of NMA to investigate the influence of the global native state topology on   
the thermal stability of proteins.

Vibrational thermal instability is a well-known issue in solid state physics. 
Since the first classical analysis of Peierls \cite{Peierls}, 
it has been recognized 
that the equilibrium with a thermal bath can dramatically influence the 
allowed topological arrangement of large geometrical structures. 
The most striking consequence of Peierls instability has concerned up 
to now low-dimensional crystals: for one and two-dimensional 
lattices the mean square
displacement of a single atom at finite temperature diverges in the
thermodynamic limit, i.e. with increasing number of atoms. 
When the displacement exceeds the order of magnitude of the lattice 
spacing, the topological arrangement of the lattice is unstable and the 
crystal becomes a liquid. For real structures, formed by a finite 
number of units and far from the
thermodynamic limit, the divergence sets a maximal stability size, which is 
negligible for one-dimensional lattices and typically
mesoscopic for two-dimensional lattices.
 
However, thermal instability is present not only in crystals but also 
in the case of structurally inhomogeneous systems, such as, glasses, 
fractals, polymers and non crystalline structures. There, 
the problem is much more complex. 
Generalizing the Peierls approach to mesoscopic disordered structures, 
we are able to apply this kind of arguments to the thermal stability of 
macromolecules. In the following we describe how this can be done in the 
case of proteins; we will predict the existence of a critical stability 
size depending on a global topological parameter (the spectral dimension) 
and compare our predictions with experimental data.

\section{THEORY}

In a recent paper \cite{BCFV} generalizing the Peierls' result, we have 
shown that a thermodynamic instability appears also in inhomogeneous 
structures and it is determined by the spectral dimension $\bar {d}$. 
The parameter $\bar{d}$ \cite{Orbach} is defined
according to the asymptotic behaviour of the density of harmonic 
oscillations at low frequencies.
More precisely, denoting by $\rho(\omega)$ the density of modes with 
frequency $\omega$, then
\begin{equation}
\rho(\omega) \sim \omega^{\bar{d}-1}
\end{equation}
for $\omega\to 0$.
The spectral dimension is the most natural 
extension of the usual Euclidean dimension $d$ to disordered 
structures as far as dynamical processes are concerned. 
It coincides with $d$ in the case of lattices, but in general, it can 
assume non-integer values between $1$ and $3$. 
The spectral dimension represents a useful measure of the
effective connectedness of geometrical structures at large scales, because 
large values of $\bar{d}$ correspond to high topological 
connectedness. 
Moreover, it characterizes not only
harmonic oscillations, but it is also related to diffusion, 
phase transitions and electrical conductivity, allowing a variety of
both experimental and numerical methods for its determination
\cite{Burioni,Saviot}.
The relevance of $\bar{d}$ in connection with the anomalous density of 
vibrational modes in proteins has also been considered in 
refs.\cite{Avraham,Karplus}.

In the case of thermal instability, we demostrated that, 
for $\bar {d} \ge 2$, the mean
square displacement $\langle r^2 \rangle$ of a structural unit (being an
atom, a molecule or a supramolecular structure according to the studied case)
of a system composed of $N$ elements, diverges in the limit $N\to\infty$.
Denoting with $T$ the temperature of the heat bath and with $K_B$ the
Boltzmann constant, the divergence is given by the asymptotic law: 
\begin{equation}
\langle r^2 \rangle \sim K_B T N^{2/\bar{d} - 1}
\label{PL}
\end{equation}
for $\bar {d}<2$. For $\bar {d} = 2$, the
mean square displacement diverges logarithmically, $\langle r^2 \rangle \sim
K_B T \ln~N$, as in the case of the Peierls' result for a two dimensional
crystal. Notice that the divergence in $\langle r^2 \rangle$
is only determined by $\bar {d}$.
Now, at any
given temperature $T$, there will exist a threshold value $N(T)$ beyond
which $\langle r^2 \rangle^{1/2}$ exceeds the typical spacing
between nearest neighbors, making the solid structure unstable. 
Therefore at large enough $N$, the solid will experience a structural 
reorganization
which can lead either to a homogeneous liquid phase at sufficiently high 
temperatures
or to a disordered 3-dimensional solid, which is homogeneous at large scales
and inhomogeneous at small scales.
In general,
the threshold values of $N$ are very small with respect to the typical
order of magnitude of macroscopic systems, being rather comparable to the size
of large complex macromolecules such as biopolymers. 

This poses an intriguing question concerning proteins. 
Indeed, to exploit their biological function
proteins must keep a specified geometrical and topological 
arrangement and cannot
afford any, even partial, geometrical large scale fluctuations as it happens,
instead, to swallen polymeric chains in a good solvent \cite{Degennes}. 
This makes thermodynamical stability problems of
crucial importance and suggests a possible correlation between the spectral 
dimension and the length of protein chains. 

The vibrational stability problem in proteins has been analysed through 
the gaussian network model (GNM), proposed in \cite{Tirion}, which
yields results in agreement with principal X-ray spectroscopy experiments. 
This approach generally considers proteins as elastic networks, whose nodes 
are the positions of the alpha-carbons (C$_{\alpha}$) in the
native structure and the interactions between nodes are assimilated to harmonic
springs. The only information required to implement the method is the knowledge
of the native structure, and only two free parameters are introduced, 
the spring
constant and interaction cutoff. The GNM is defined
by the quadratic Hamiltonian 
\begin{equation}
H = \sum_i^{N} \frac{{\bf p}_i^2}{2 m} +
\frac{\gamma}{2}\sum_{ij} \Delta_{ij}(\delta{\bf r}_i - \delta{\bf r}_j)^2
\label{eq:GNM}
\end{equation}
where the first term is the kinetic energy of the system,  
$\gamma$ being the strength of the springs that are assumed 
homogeneous, ${\bf R}_i$ and $\delta {\bf r}_i$ indicating 
the equilibrium position and the displacement with respect to ${\bf R}_i$
of the $i$-th  C$_{\alpha}$ atoms. The model is eventually defined by the 
contact
matrix ${\Delta}$ with
entries: $\Delta_{ij}=1$ if the distance $|{\bf R}_i - {\bf R}_j|$
between two C$_{\alpha}$'s, in the native conformation,
is below the cutoff $R_0 = 0.6$ \AA, while is $0$ otherwise. 

The harmonic spectrum for each structure is given by the set of 
eigenvalues $\{\omega_1,...,\omega_N\}$ of the
Kirchhoff matrix (or valency-adjaciency matrix)
$\Gamma_{ij} = - \Delta_{ij} + \delta_{ij} \sum_{l\neq i} \Delta_{il}$,

Notice that
the first eigenvalue $\omega_1$ vanishes and corresponds to the constant 
eigenvector,
describing the trivial uniform translation.

The comparison between experimental data and GNM results is obtained via the
X-ray crystallographyc B-factors, measuring the mean square
fluctuation of C$_\alpha$ atoms around native positions

$$
B_i(T) = \frac{8\pi^2}{3}\langle \delta{\bf r}_i \delta{\bf r}_i \rangle
$$

In the GNM approximation, these can be expressed in terms of the 
diagonal part of the inverse of $\Gamma$~\cite{Bfact}:  

$$
\langle \delta{\bf r}_i \delta{\bf r}_j \rangle = \frac{3 K_B T}{\gamma}
[\Gamma^{-1}]_{ij}
$$

The comparison with crystallographyc data is crucial for setting 
the correct values of the free parameters $R_0$ end $\gamma$ of the GNM 
(see. Methods and Results).

\section{METHODS AND RESULTS}

We present
now the harmonic analysis of the GNM performed over the dataset 
of protein native
structures with different sizes listed in Table 1. and downloaded from the
Brookheaven Protein Data Bank. We selected our representative statistical
sample according to the following criteria: 
first, we only considered proteins with a stable large 
scale geometry.
This excludes multiple domains proteins, where the domains can move
independently giving rise to large geometrical fluctuations. 
Moreover, we selected proteins not binded to fragments of
DNA, RNA or other substrates because such structures cannot be described 
with sufficient accuracy in terms of simple
harmonic model with only two effective parameters. 

Finally, we choose proteins covering uniformly a wide length interval 
ranging from $100$ to $3600$ to test our prediction.

The diagonalization of the Kirchhoff matrix $\Gamma$ to obtain 
its eigenvalues $\{\omega_1^2,...,\omega_n^2\}$ and eigenvectors has been 
performed with the standard numerical packages \cite{Recipes}. The knowledge
of the eigenvectors (eigenmodes) allows to compute the GNM B-factors:
$$
B_i(T) = \frac{8\pi^2 K_B T}{\gamma} \sum_{k} \frac{|u_i(k)|^2}{\omega_k^2}
$$
where $i$ is the residue index, the sum runs over all non-zero 
frequencies  and 
${\bf u(k)}_i$ indicates the $i$-th component of the $k$-th eigenmode.   

\begin{figure}
\includegraphics[clip=true,width=\columnwidth,keepaspectratio]
{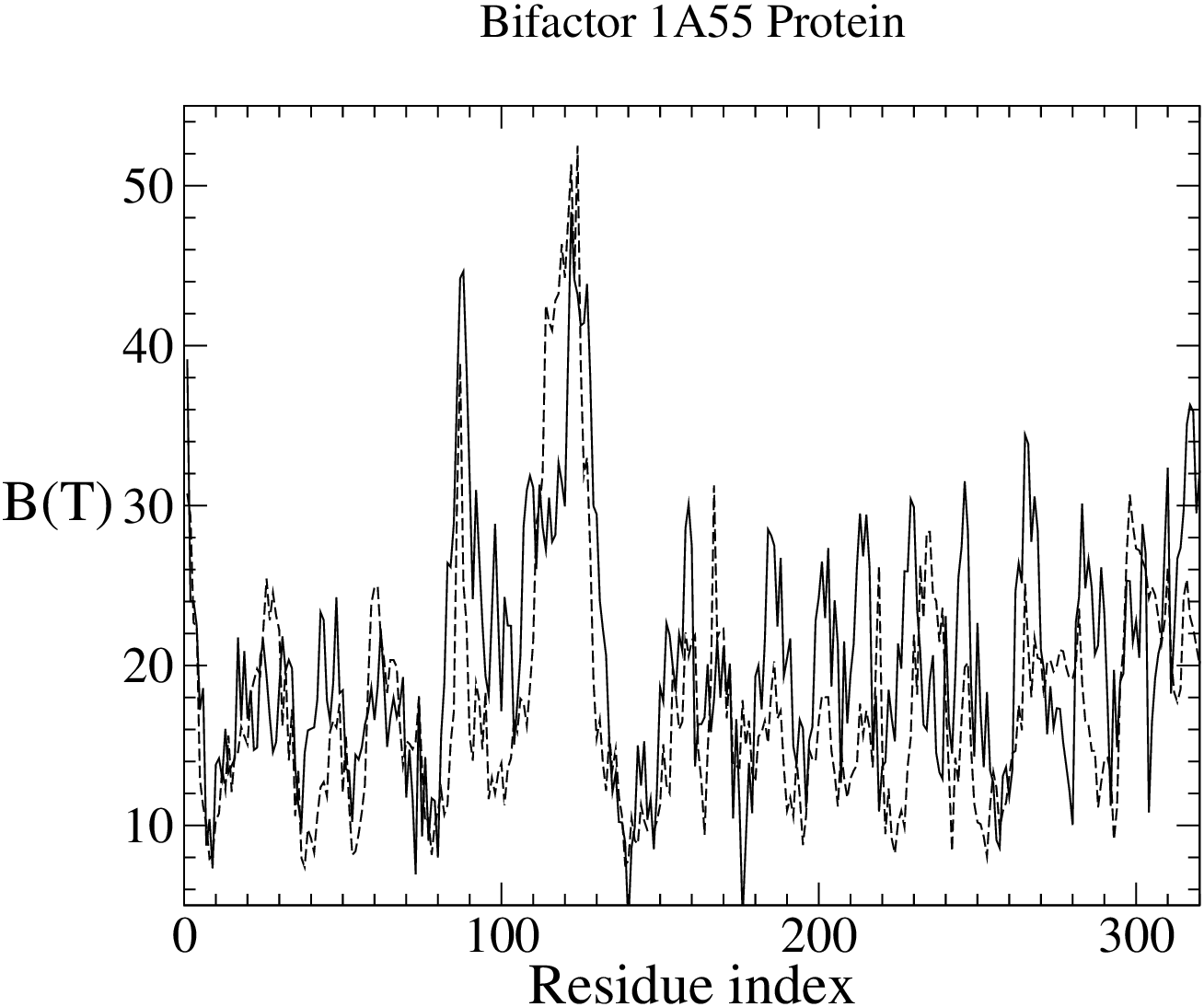}
\caption{\label{bfact} Comparison
of experimental B-factors and mean square fluctuations of Ca yielded by
GNM with a cutoff $R_0 = 6~\AA$ for the protein $1A55$ with 321 residues.
Solid line refers to crystallographic data, while dashed refers to GNM
approximation. Since, there is a reasonable overlap between flexible
regions of analysed proteins predicted by GNM and experimental indications,
this is a good check for the correctness of the cutoff choiche.}
\end{figure}

When generating the contact matrix $\Delta$, the value of the interaction 
cutoff $R_0$ has been chosen in order to have agreement between the 
experimental and the GNM B-factors for each proteins, as showed
in Fig.~\ref{bfact}. 
We found that the average value 
$R_0 = 6~\AA$ was a consistent choice for the whole set of proteins.

The purpose of our analysis is basically:

a) investigating
whether there exists a correlation between the spectral dimension of native
structures and the length of natural occurring proteins;

b) in the
case of positive result for point a), verifying whether the correlation is
connected with equation (\ref{PL}). 

This expression establishes a rather strong relation between 
the spectral dimension and the
maximum size $N_{max}$ of a protein. Since the stability is
supposed to fail when the fluctuation $\langle r^2 \rangle^{1/2}$
becomes of the same order of magnitude of the mean distance between 
non consecutive aminoacids ($6$~\AA), one has 

\begin{equation}
\frac{1}{\ln(N_{max})} = a(2/\bar{d} -1)
\label{div}
\end{equation}

The proportionality constant $a$ in (\ref{div}) depends on the mean 
aminoacid spacing, on the spring elastic constant g and temperature T. 
However, this dependece is very weak
(i.e.only logarithmic). This allows for a comparison of different proteins
without the computation of the specific parameters. 

Actually,
equation (\ref{div}), being based uniquely on thermodynamics stability, 
can be regarded as an upper bound prediction only. 

In Figure~2, we plot the harmonic spectrum, obtained within the GNM,
for three proteins with
sizes, small, medium and large, respectively. The low frequency regions clearly
exhibit a power law behavior whose exponent $\beta$ is directly related to
the spectral dimension via formula $ \bar {d} = 2/\beta$.

Figure~3 summarises the results of our analysis.
We plot the $1/\ln(N)$ versus the computed slope from the spectra.
The choice of these variables is suggested by relation (\ref{div}):
indeed, if eqn.(3) holds,
we should obtain a straight line crossing the origin.
 
As matter of fact, our data
are very well fitted by a straight line, with linear regression coefficient
$0.76$, but described by the equation
\begin{equation}
\frac{1}{\ln(N)} = a(2/\bar{d} -1) + b
\label{parb}
\end{equation}
with parameters $a=0.133$ and $b = 1.134$.

\section{Discussion and Conclusions}

The neat result expressed by (\ref{parb}) deserves several comments. First,
this equation is in agreement with
the upper bound we obtained  Eq.~\ref{div}, supporting the relevance of 
topological thermal
instability in constraining the protein geometry. More important, not only the
upper bound is satisfied, but the experimental points lie on a straight line
parallel to the upper bound one Eq.~(\ref{div}). 
This suggests a more fundamental role
of topological stability: in some sense, the protein tends to arrange
topologically in such a way to reach the minimum value compatible with 
stability constraints. 
In other words, for any fixed length, it tends to the most swollen
state which remains stable with respect to thermal fluctuations.
  
A very
interesting point is the meaning of the parameter $b$ which is 0 in 
Eq.~(\ref{div}). Its
positive value could have different motivations, but its universal nature (it
is a ``protein-independent'' global shift) must be due to a very general
mechanism. A rather obvious reason is the contribution of anharmonic
interactions at finite temperatures; a more intriguing one could be an
effective longer range interaction due to the presence of bound water molecules
around the external aminoacids, which could change the effective form of the
interaction matrix $\Lambda$. This hypothesis is also suggested by the
physical interpretation of $b$ as an anomalous dimension exponent,
typically related to a renormalized interaction \cite{Golden}  

However, the most intriguing evidence relies on the regression coefficient
Independently of the physical origin of $b$, its high value strongly supports
the existence of a thermodynamic stability threshold, dependent on the 
topology of the folded state, for the size of proteins.



\begin{figure}
\includegraphics[clip=true,width=\columnwidth,keepaspectratio]
{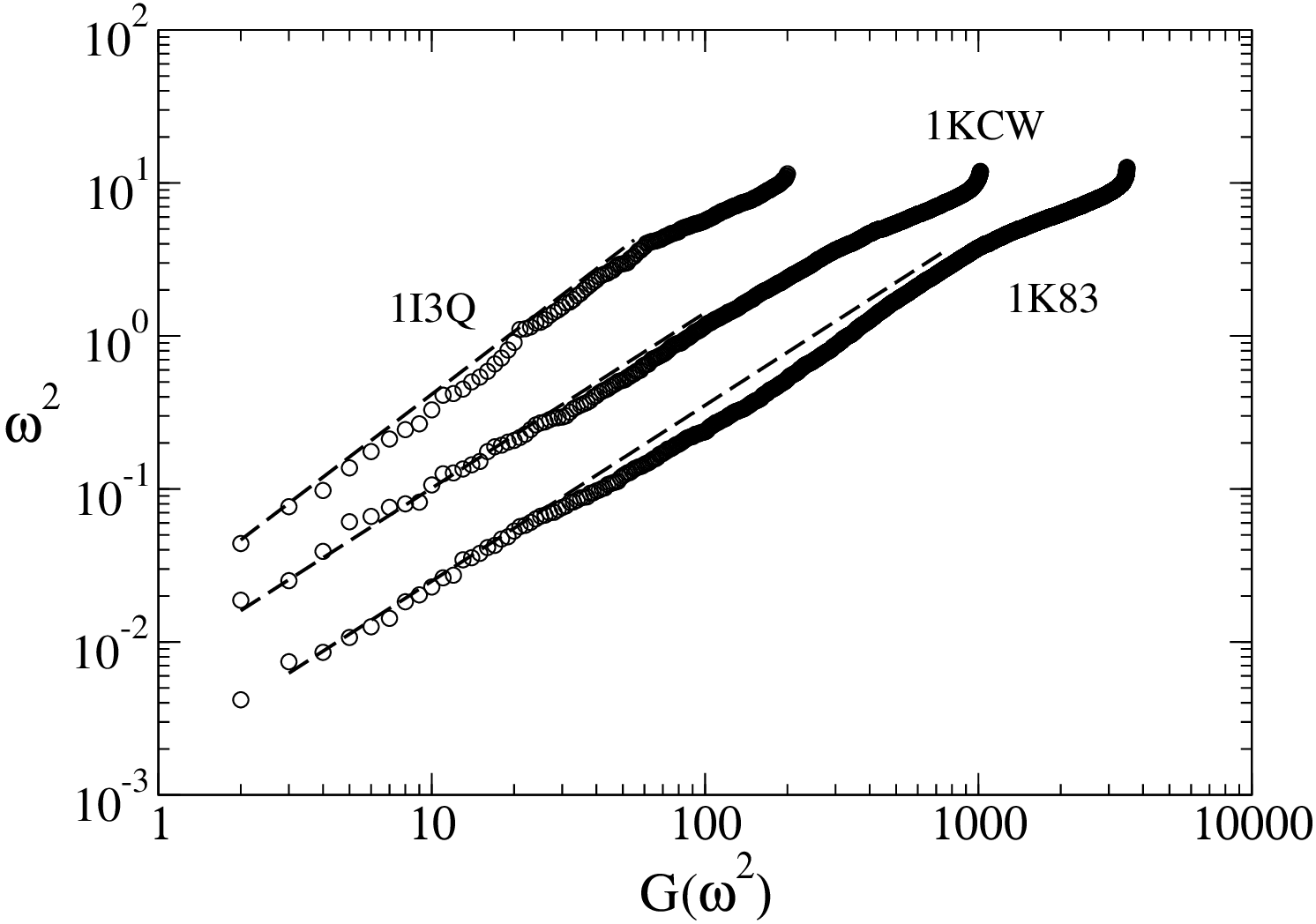}
\caption{Log-log plot
of the harmonic spectrum from GNM for three proteins of different sizes, 1I3Q
(N=200), 1KCW (N=1017) and 1K83 (N=3494). On vertical axis, we report the
eigenvalues $\omega^2$ of the matrix $\Gamma$, and on the orizontal, the 
number of eigenvalues below a given value $\omega^2$, i.e. the
integrated density of modes. Dashed lines are the best-fits, 
whose slope $\beta$, 
is related to the spectral dimension, $\beta = 2/\bar d$}
\end{figure}

\begin{figure} 
\includegraphics[clip=true,width=\columnwidth,keepaspectratio]
{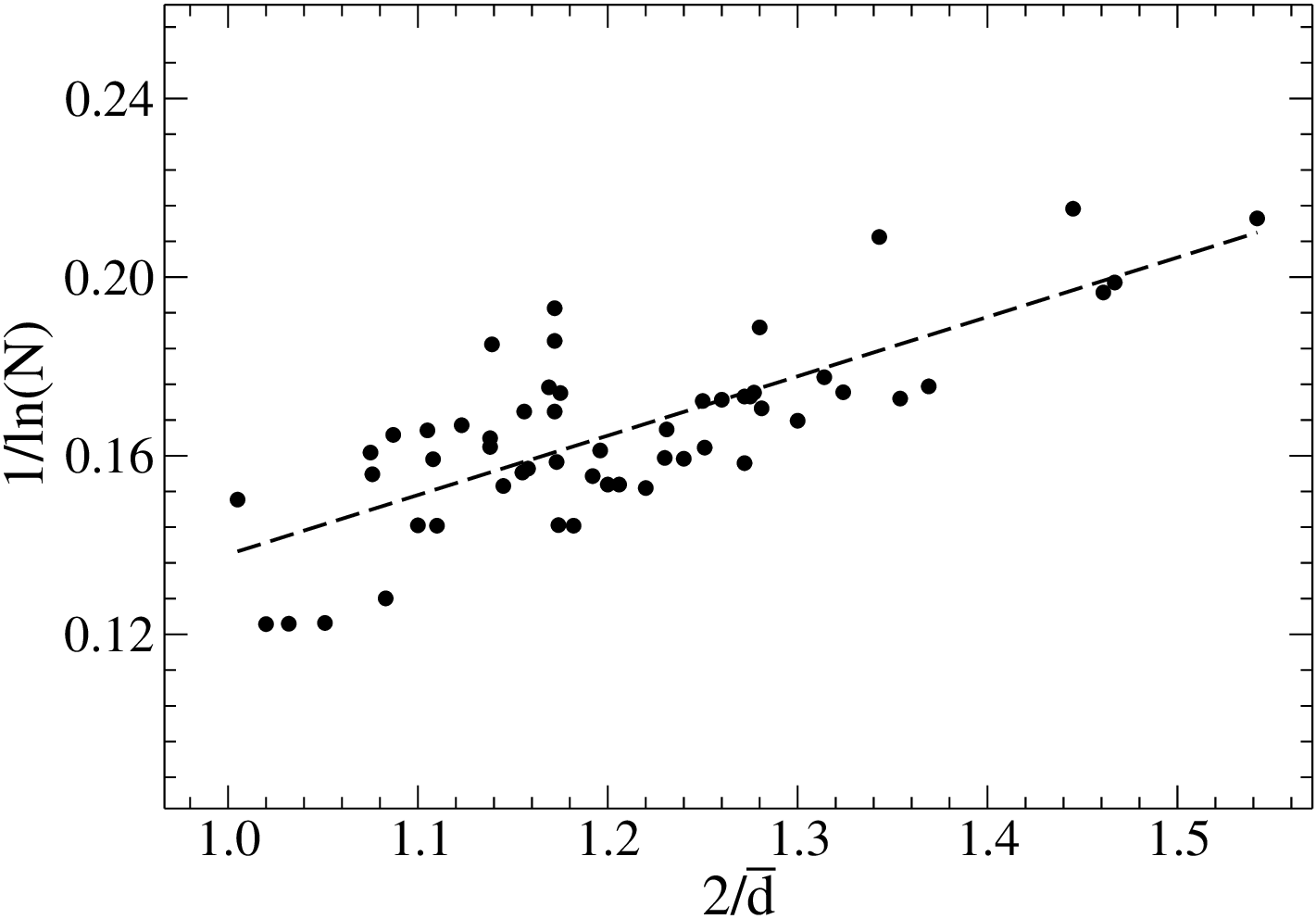}
\caption{Linear plot showing the dependence of the spectral dimension on
protein size. The line, indicating the behaviour (4), is a best fit with a
correlation coefficient 0.76 .}
\end{figure}

\begin{table}
\begin{ruledtabular}
\begin{tabular}{|l|l|l||l|l|l|}
PDB code & Length & $\overline{d}$ & 
PDB code & Length & $\overline{d}$   \\
\hline\hline
9RNT & 104 & 1.38 & 1DY4 & 434 & 1.84 \\
1THO & 109 & 1.30 & 1BU8 & 446 & 1.76 \\
1CCR & 111 & 1.49 & 1AUK & 480 & 1.76 \\ 
1BVC & 153 & 1.36 & 1AC5 & 483 & 1.60 \\
110L & 162 & 1.37 & 1CPU & 495 & 1.67 \\
1G12 & 167 & 1.61 & 1A65 & 504 & 1.86 \\
1AMM & 174 & 1.55 & 1SOM & 528 & 1.63 \\
1IQQ & 200 & 1.57 & 1E3Q & 532 & 1.61 \\
1B0F & 218 & 1.67 & 1CRL & 534 & 1.81 \\ 
1AE5 & 223 & 1.66 & 1AKN & 547 & 1.71 \\ 
1A06 & 279 & 1.52 & 1DIY & 553 & 1.58 \\  
1A48 & 298 & 1.46 & 1CF3 & 582 & 1.73 \\
1A3H & 300 & 1.71 & 1EX1 & 602 & 1.73 \\  
16VP & 311 & 1.51 & 1A14 & 612 & 1.86 \\  
1A5Z & 312 & 1.57 & 1MZ5 & 622 & 1.68 \\  
1A1S & 313 & 1.70 & 1CB8 & 674 & 1.66 \\
1A40 & 321 & 1.57 & 1HMU & 674 & 1.67 \\  
1A54 & 321 & 1.57 & 1A47 & 683 & 1.75 \\  
137L & 326 & 1.48 & 1DMT & 696 & 1.64 \\  
1A20 & 329 & 1.59 & 1A4G & 780 & 1.99  \\ 
1A0I & 332 & 1.60 & 1HTY & 1014 & 1.70 \\  
1A26 & 351 & 1.56 & 1KCW & 1017 & 1.82 \\  
8JDW & 360 & 1.71 & APP1 & 1021 & 1.69 \\  
1BVW & 360 & 1.73 & 1B0P & 2462 & 1.80 \\  
7ODC & 387 & 1.54 & 1KEK & 2462 & 1.94 \\  
1A39 & 401 & 1.78 & 1K83 & 3494 & 1.90 \\  
16PK & 415 & 1.63 & 1I3Q & 3542 & 1.94 \\  
1A2N & 418 & 1.81 & 1I50 & 3558 & 1.96 \\  
\end{tabular}
\caption{\label{tab:proteins}
List of processed native protein structures from Brookheaven PDB,
with their lenght and correspondig spectral dimension
estimated by GNM approach.}
\end{ruledtabular}
\end{table}

\end{document}